\documentclass[12pt, onecolumn,twoside]{IEEEtran}
\usepackage{cite}
\usepackage[pdftex]{graphicx}
\usepackage{setspace}
\usepackage{amsmath}
 \usepackage{amssymb}
\usepackage{amsfonts}
\usepackage{subfig}
\usepackage{textcomp}
\usepackage{wrapfig}
\usepackage{lipsum}
\usepackage{multicols}
\usepackage{amsmath,graphicx}
\def\BibTeX{{\rm B\kern-.05em{\sc i\kern-.025em b}\kern-.08em
    T\kern-.1667em\lower.7ex\hbox{E}\kern-.125emX}}

\begin{document}

\title{A note on dual demodulator continuous transmission frequency modulation technique}

\author{\normalsize{ Kapil Dev Tyagi,
Arun Kumar and Rajendar Bahl
\\ kapil.tyagi@jiit.ac.in, \{arunkm, rbahl\}@care.iitd.ac.in
\\CARE, Indian Institute of Technology, Delhi.}}
{}

\maketitle

\begin{abstract}
The range resolution in conventional continuous time 
frequency modulation (CTFM) is inversely proportional to the signal bandwidth. The
dual-demodulator continuous time frequency modulation (DD-CTFM) processing technique was proposed by Gough et al \cite{0.18} as a method 
to increase the range resolution by making the output of DD-CTFM truly continuous.
However, it has been found that in practice the range resolution is still limited by the signal bandwidth.
The limitation of DD-CTFM has been explained using simulations and mathematically in this paper.


\end{abstract}


\section{Introduction}
According to Gough et al \cite{0.18}, ``by using a dual-demodulator continuous time frequency modulation (DD-CTFM) system the demodulated 
signal can be made continuous, resulting in the complete elimination of the blind time and in a range resolution as good as two wavelengths 
of the mean frequency''. 
However, in practice DD-CTFM technique does not offer significant improvement in range resolution.
Actually, the range resolution improvement is only 10 to 20 \% depending on the blind time duration over a conventional CTFM technique. 
The range resolution of CTFM technique is $\frac{C}{2B}$, where $C$ is the speed of the wave and $B$ is the bandwidth of the signal \cite{1.1L}. 
The reason for the limitation of DD-CTFM technique is the phase 
mismatch at the boundaries of the concatenated segments from the two channels in the difference frequency signal in each sweep cycle.
The phase mismatch problem makes the claim of Gough et al \cite{0.18} to be incorrect which states that ``we have managed to eliminate
the blind time associated with conventional CTFM sonars, thus making the output truly continuous''.
This issue was not addressed in \cite{0.18} and is explained in this paper.

The next section discusses the basic principle of DD-CTFM technique.
Section III describes the issues in obtaining higher range resolution using DD-CTFM. 
The final section concludes this paper.

\section{Basic Principle of DD-CTFM}
\begin{figure}[htb]
 \centering
 \includegraphics[width=155mm,height=93mm]{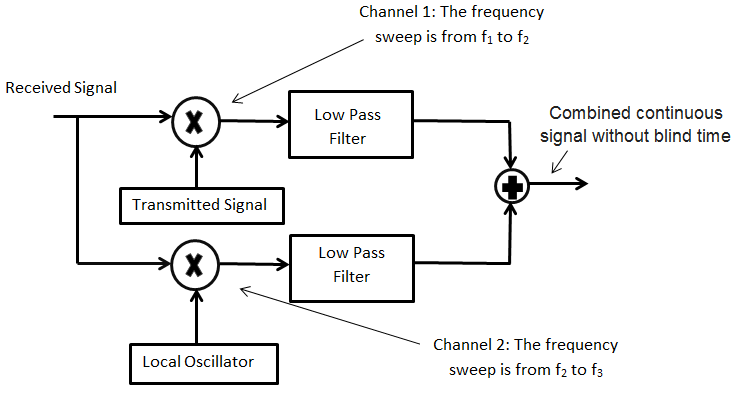}

 \caption{Receiver processing for DD-CTFM technique.}
 \label{fig5.3}
\end{figure}

The problem of blind time limits the observation time and hence 
the range resolution in CTFM \cite{0.18}.
According to Gough et al \cite{0.18}, DD-CTFM receiver processing eliminates the blind time and thereby increases 
the observation time leading to high resolution in frequency and hence range. 
As proposed in Fig \ref{fig5.3}, the output of DD-CTFM processing should ideally consist of continuous sinusoids, whose
frequencies are proportional to the range of the reflection interfaces.  
In the output of DD-CTFM processing, continuity of sinusoids across the sweep 
  periods is achieved by
  extending the transmitted
  signal by a duration equal to the maximum blind time using a local oscillator in the second channel as shown in Fig \ref{fig5.3}.
  The local oscillator generates a linear frequency in phase continuity with the transmitted signal with identical 
  slope same as the transmitted signal. 
  The frequency of the local oscillator 
  increases from $f_{2}$ to $f_{3}$.
  Thus, the receiver has two channels, with 
  the processing in channel 1 being the same as in CTFM technique \cite{1.1}. 
  In channel 1 the 
  received signal is multiplied with the transmitted signal and low pass filtered while in channel 2, the received signal is multiplied 
  with the signal from the local oscillator and low pass filtered. 
  To obtain a continuous waveform, the outputs of both channels are added.
  Now, the added low pass filtered signals from both the channels
  can be ideally expected to be continuous with no blind time. 
  Hence, one can ideally expect to obtain any desired frequency resolution, 
  as the observation time can be extended to any duration.

\section{ Issues in DD-CTFM }

\begin{figure}[!h]
 \centering
 \includegraphics[width=155mm,height=88mm]{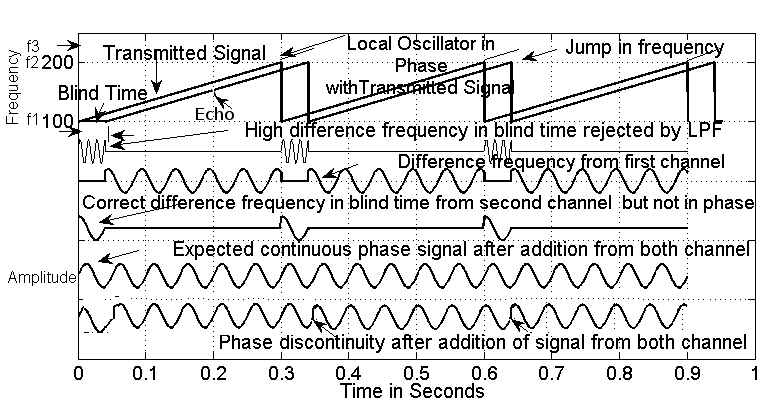}
 \caption{Time-frequency plot and DD-CTFM processing: difference between desired waveform (2$^{nd}$ from bottom) and actual waveform
 (bottom).}
 \label{fig5.5}
\end{figure}

In doing careful analysis of the DD-CTFM technique, it is found that there is a problem of phase discontinuity in the output signal. 
The output signal of DD-CTFM processing should ideally contain a sinusoidal waveform without any phase discontinuity,
corresponding to each echo.
However, there will be phase discontinuity in the concatenated sinusoids from the two channels as 
illustrated in Fig. \ref{fig5.5}. 

The time-frequency plot and difference frequency signal waveforms are shown in Fig. \ref{fig5.5}.
The upper part of Fig. \ref{fig5.5} shows the time-frequency plots of the transmitted and 
received signals for a single target. The lower part of Fig. \ref{fig5.5} shows the incorrect jump frequency (that does 
not contain the range information) and
correct difference frequency signals respectively. 
The correct difference frequency
signal refers to the desired signal corresponding to the actual delay. 
The jump difference frequency signal is incorrect as it contains wrong information of the delay. As shown
in Fig. \ref{fig5.5}, the correct difference frequency signal is obtained from a segment of the
total cycle duration only from channel 1. 
In the remaining part of each cycle just after the
frequency jumps in the transmitted signal, the difference frequency signal becomes incorrect in channel 1. 
The duration for which we get the incorrect difference frequency signal is
known as blind time and it depends on the target range. 
The blind time limits the total observation time of the correct frequency and hence limits
the frequency resolution. To get the correct difference frequency signal during the blind time, a second channel with a 
local oscillator was proposed by Gough et al \cite{0.18}. The desired added signal from both the channels is shown in Fig. \ref{fig5.5} by the second last waveform, 
which does not contain any phase discontinuity. 
However, the actual concatenated sinusoids shown by the bottom waveform contains the phase discontinuity, which limits the observation time
and hence the range resolution.

\subsubsection*{\textbf{Mathematical explanation of phase discontinuity in the added signal}}

\hspace{2mm}Let the transmitted signal $x(t)$ be a linear frequency modulated signal (up-chirp) represented as:
\begin{equation}\label{e7.2}
 x(t)=cos \Big(2\pi (f_{1}t+\frac{\mu t^{2}}{2} ) \Big)
\end{equation}
where $f_{1}$ represents the starting frequency of the LFM signal and $\mu$ is the sweep rate. Let $\phi_{x}(t)$ be the instantaneous 
phase of the transmitted signal given as: 
\begin{equation}\label{px}
 \phi_{x}(t)=2\pi (f_{1}t+\frac{\mu t^{2}}{2} ).
\end{equation}
Let the received signal $y(t)$ be 
a delayed version of the transmitted signal given as:
\begin{equation}\label{e7.3}
 y(t)=cos \Big(2\pi (f_{1}(t-\tau)+\frac{\mu}{2} (t-\tau)^{2}) \Big)
\end{equation}
where $\tau$ is the time delay of the received signal. 
The instantaneous phase of $y(t)$ be expressed as:
\begin{equation}\label{py}
\phi_{y}(t)= 2\pi (f_{1}(t-\tau)+\frac{\mu}{2} (t-\tau)^{2})
\end{equation}
where $\phi_{y}(t)$ represents the phase of $y(t)$ at time $t$.

The transmitted and received signals are multiplied and 
low pass filtered in channel 1 as shown in Fig. \ref{fig5.3}. To get the ideally desired signal 
corresponding to a range cell during the blind time, a local oscillator is used.
According to Gough et al \cite{0.18} if the local oscillator is in phase continuity with the transmitted signal and with 
the same slope, we 
should get a continuous sinusoidal signal after the addition of the two channels. 
Let the local oscillator signal which is
in phase continuity with the transmitted signal be given as:
\begin{equation}\label{e7.4}
z(t)=cos \Big(2\pi (f_{2}t+\frac{\mu}{2} t^{2})+\phi_{I} \Big)
\end{equation}
where $f_{2}$ is the starting frequency of the local oscillator, that is equal to the final frequency of the transmitted 
chirp signal. The initial phase $\phi_{I}$ of the local oscillator is equal to the phase 
of the transmitted signal at the instant when the transmitted signal frequency reaches its final value. 
The initial phase $\phi_{I}$ is calculated using (\ref{px}) as:
\begin{equation}
\phi_{I}= 2\pi (f_{1}T+\frac{\mu}{2} T^{2}).
\end{equation}
Let the phase of $z(t)$ be denoted by $\phi_{l}(t)$. It is given as:
\begin{equation}\label{pl}
\phi_{l}(t)=2\pi (f_{2}t+\frac{\mu}{2} t^{2})+\phi_{I}.
\end{equation}
The phase of the low pass filtered signal at the output of channel 1 is calculated as:
\begin{equation}\label{pc1}
\phi_{f1}(t)= \phi_{x}(t)-\phi_{y}(t).
\end{equation}
The phase of the low pass filtered signal at the output of channel 2 is calculated as:
\begin{equation}\label{pc2}
\phi_{f2}(t)= \phi_{l}(t)-\phi_{y}(t).
\end{equation}

Let the starting frequency $f_{1}$, the final frequency $f_{2}$ and the duration $T$ of the chirp signal be taken as
100 Hz, 200 Hz and 300 ms respectively.
Let the delay $\tau$ in the received signal be 93 ms. The value of $\mu$ for this case is 333.3 Hz/sec. The phase of the transmitted signal at 
$t=300$ ms  using (\ref{px}) is  
$(90 \pi)_{mod 2\pi}=0$ radians. The phase of the received signal at this instant using (\ref{py}) is $(55.68)_{mod 2\pi}=1.68\pi$ radians. 
So, the phase of the low pass filtered signal using (\ref{pc1})
at $t=300$ ms from channel 1 is $(34.32 \pi)_{mod 2\pi}=0.32\pi$ radians.
At $t=300$ ms,
the transmitted signal will jump back to its starting frequency of 100 Hz for the next cycle. 
Now, the desired signal during the blind time is provided by channel 2. Using (\ref{py}) and (\ref{pl}), 
the phase of the received signal $\phi_{y}$
and local oscillator signal $\phi_{l}$ at $t=300$ ms are $(55.68)_{mod 2\pi}=1.68\pi$ and $(90 \pi)_{mod 2\pi}=0$ radians respectively. So, using (\ref{pc2}) the starting 
phase of the low pass filtered 
signal from channel 2 is $(34.32 \pi)_{mod 2\pi}=0.32\pi$ radians. Thus, the phase of signals from the two channels is continuous without any
discontinuity at $t=300$ ms. 

At $t=393$ ms, the desired signal is obtained from channel 2. The phase 
of the local oscillator $\phi_{l}$ and the received signal $\phi_{y}$ at $t=393$ ms using (\ref{pl}) and (\ref{px}), 
are $(130.08 \pi)_{mod 2\pi}=0.08\pi$ and $(90 \pi)_{mod 2\pi}=0$ radians 
respectively. 
So, the phase of the low pass filtered signal 
from channel 2 at the end of the first cycle using (\ref{pc2}) is $(40.08 \pi)_{mod 2\pi}=0.08\pi$ radians.
The desired signal just after $t=393$ ms is again obtained from channel 1. At $t=393$ ms in the next cycle, the phase
of the transmitted $\phi_{x}$ and the received signals $\phi_{y}$ using (\ref{px}) and (\ref{py}), are $(21.48 \pi)_{mod 2\pi}=1.48\pi$ and $(90 \pi)_{mod 2\pi}=0$ radians 
respectively. The starting phase of 
channel 1 in the second cycle using (\ref{pc1}) is $(-68.52 \pi)_{mod 2\pi}=-0.52\pi$ radians that is very different from the previous phase of $(40.08 \pi)_{mod 2\pi}=0.08\pi$. 
Thus, the added difference frequency signal from the two channels 
contains a strong phase discontinuity after $t=393$ ms.
Thus, the observation duration of the sinusoidal signal corresponding to an echo
is increased only by a duration equal to blind time, which is a fraction of total chirp duration. So, one cannot increase the 
observation duration to any desired value as claimed in \cite{0.18}.
The expressions for the phase of the two channels at various instants with a typical example explained above
are summarized in table \ref{T5MDAA} to demonstrate the phase discontinuity issue.

\begin{table}[h]
\begin{tabular}{ | l | l | l | }

\hline
   Parameter          & Expression       & Example\\ \hline
   Starting frequency  &      $f_{1}$           &  100 Hz  \\ \hline  
   Final frequency of $x(t)$  &      $f_{2}$           &  200 Hz  \\ \hline   
   Sweep duration       &        $T$           &  300 ms \\ \hline
   Delay in $y(t)$       &       $\tau$            &  93 ms \\ \hline
   Sweep rate         & $\dfrac{(f_{2}-f_{1})}{T}$              &    333.3 Hz/sec.           \\ \hline           
   Phase of $x(t)$ at $t=T$ &  $\phi_{x}(t)=2\pi (f_{1}T+\dfrac{\mu}{2} T^{2})$   &   90$\pi$            \\ \hline
   Phase of $y(t)$ at $t=T$ & $ \phi_{y}(t)= 2\pi (f_{1}(T-\tau)+\dfrac{\mu}{2} (T-\tau)^{2}) $  &  55.68$\pi$ \\    \hline        
   Phase of $z(t)$ at $t=T$  & $\phi_{l}(t)= \phi_{i} $                                     &  90$\pi$     \\ \hline
  
    Phase of $x(t)$          & $\phi_{x}(t)=2\pi (f_{1}\tau+\dfrac{\mu}{2} \tau^{2})$                 &  40.08$\pi$        \\ 
    at $t=T+\tau$ &                                                   &  \\ \hline 
   
    Phase of $y(t)$ &  $ \phi_{y(t)}= 2\pi (f_{1}T+\dfrac{\mu}{2} T^{2}) $      &   90$\pi$  \\ 
    at $t=T+\tau$        &                             &        \\ \hline

    Phase of $z(t)$  &    $\phi_{l}(t)=2\pi (f_{2}\tau+\dfrac{\mu}{2} \tau^{2})+\phi_{i} $      &   130.08$\pi$  \\  
    at $t=T+\tau$ &                                  &     \\ \hline 
   
   Phase of  &   $\phi_{f2}(t)= \phi_{x}(t)-\phi_{y}(t)$    & 34.32$\pi$             \\  
   channel 1 at $t=T$ &                                  &                            \\ \hline 
   
   Phase of &  $\phi_{f2}(t)= \phi_{l}(t)-\phi_{y}(t)$   &          34.32$\pi$  \\  
   channel 2 at $t=T$ &                                 &                   \\ \hline 
   
   Phase of channel 1 &   $\phi_{f2}(t)= \phi_{x}(t)-\phi_{y}(t)$    &                    -49.92$\pi$  \\  
    at $t=T+\tau$ &                            &                            \\ \hline 
   
   Phase of channel 2 &  $\phi_{f2}(t)= \phi_{l}(t)-\phi_{y}(t)$   &                    40.08$\pi$  \\  
    at $t=T+\tau$ &                                 &                   \\ \hline 
   
\end{tabular}
\caption{Parameters for calculation of phase mismatch in DD-CTFM processing.}
  \label{T5MDAA}
\end{table}

\subsubsection*{\textbf{Simulation study of phase discontinuity in the added signal}}
\begin{figure}[!h]
 \centering
 \includegraphics[scale=0.8]{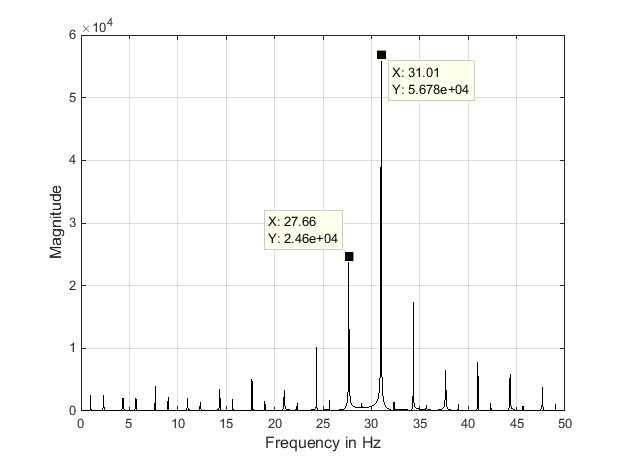}
 \caption{DFT magnitude plot of the output signal in DD-CTFM processing with phase discontinuity problem.}
 \label{fig5.26}
 \end{figure}

\hspace{2mm} For the simulation study the parameters starting frequency $f_{1}$, final frequency $f_{2}$ and duration $T$ of the chirp 
signal are taken to be 100 Hz, 200 Hz and 300 ms 
respectively. The parameters starting frequency $f_{2}$, final frequency $f_{3}$ and duration $T$ of the chirp signal generated 
using local oscillator are taken to be 200 Hz, 240 Hz and 120 ms respectively. 
The number of sweep cycles are taken to be 12, so the output signal duration is 3.6 seconds.
The maximum duration of blind time is assumed to be 120 ms.
The initial phase of the transmitted chirp signal is taken as 0. The initial phase of the local oscillator is taken 
to be same as the phase of the transmitted chirp signal at 300 ms. Let the received signal consist of a delayed version of the 
transmitted signal. Let the delay in the received signal be 96 ms.
So, the DD-CTFM output frequency in this simulation 
should ideally be 32 Hz.
However, it is estimated as 31.01 Hz as observed from Fig. \ref{fig5.26} for this simulated case.
Due to phase discontinuities in every
sweep of the chirp signal, the observation time is still limited just as in case of CTFM processing and hence there will be an error in
the estimate of the correct difference frequency.
Thus, the range resolution improvement in DD-CTFM technique  
is only slightly better depending upon the blind time duration compared to $\frac{C}{2\bigtriangleup f}$ of CTFM technique. 
This improvement is negligible compared to the claim made in \cite{0.18} which states that ``the output will be truly continuous 
resulting in the complete elimination of the blind time and in a range resolution as good as two wavelengths 
of the mean frequency''.

\subsubsection*{Sidelobe artifact issue}
There is another issue of sidelobe artifacts in the DFT magnitude of the
DD-CTFM output signal. 
The artifacts are located at integer
multiples of $\dfrac{1}{T}$ Hz around the estimated frequency in the form of peaks at several bins in DFT of the DD-CTFM output signal. 
For the simulation example, the side lobe artifacts exist at integer multiples of $3.33$ Hz as shown in Fig. \ref{fig5.26}, 
from $31.01$ Hz . 
The side lobe artifacts are only 2.3 times i.e., 7.26 dB down the main peak as shown in Fig. \ref{fig5.26}.
Thus, another weaker target echo with amplitude up to 7.26 dB below the main echo magnitude,
will not be detected. 

\section{Conclusion}
The issues in DD-CTFM technique proposed by Gough et al \cite{0.18} are presented. 
The mathematical analysis and simulation example demonstrate the incorrect inference of DD-CTFM technique which states that 
which states that ``we have managed to eliminate
the blind time associated with conventional CTFM sonars, thus making the output truly continuous'' \cite{0.18}.


\nocite{*}
\bibliographystyle{IEEE}

\end{document}